\documentclass[10pt]{iopart}
\usepackage{iopams} 
\usepackage{graphicx}
 
\begin{document}

\title[Fractional dynamic symmetries and the ground state properties of nuclei]
{Fractional dynamic symmetries and the ground state properties of nuclei}

\author{Richard Herrmann}

\address{
GigaHedron, Farnweg 71, D-63225 Langen, Germany
}
\ead{herrmann@gigahedron.com}
\begin{abstract}
Based on the Riemann- and Caputo definition of the fractional derivative we use the fractional
extensions of the standard rotation group $SO(3)$ to construct a higher dimensional representation
of a fractional rotation group with mixed derivative types. 
An extended symmetric rotor model is derived, which predicts the sequence of magic proton and
neutron numbers accurately. The ground state properties of
nuclei are correctly reproduced within the framework of this model.
\end{abstract}

%
%

\section{Introduction}
The experimental evidence for discontinuities in the sequence of atomic masses, 
$\alpha-$ and $\beta$- decay systematics  and
binding energies of nuclei suggests the existence of a set of magic proton and neutron numbers,
which can be described  successfully by single particle shell models with a heuristic spin-orbit term \cite{gop},
\cite{hax}. The most prominent representative is the phenomenological Nilsson model \cite{nie}
with an anisotropic oscillator potential:
\begin{equation} 
\label{nilsson}
V(x_i) = \sum_{i=1}^3 {1 \over 2} m \omega_i^2 x_i^2 - \hbar \omega_0 \kappa(2  \vec{l}\vec{s}+ \mu l^2) 
\end{equation} 
Although these models are flexible enough to reproduce the experimental results, they lack a deeper theoretical
justification, which becomes obvious, when extrapolating the parameters $\kappa$, $\mu$, which
determine the strength of the spin orbit and $l^2$ term  to the region of
superheavy elements \cite{hof}. 

Hence it seems tempting to describe the experimental data with alternative methods. Typical
examples are relativistic mean field theories \cite{rufa},\cite{ben}, where nucleons are described by the
Dirac-equation and the interaction is mediated by mesons. Although  a spin orbit force is obsolete in 
these models, different parametrizations predict different shell closures \cite{rutz},\cite{kruppa}. Therefore
the problem of a theoretical foundation of magic numbers remains an open question since Elsasser \cite{elsasser}
raised the problem 75 years ago.

A  fundamental understanding of magic numbers for protons and neutrons
may be achieved if the underlying corresponding symmetry of the nuclear many body system is determined. 
Therefore a group theoretical approach seems appropriate.
 
Group theoretical methods have been successfully applied to problems in nuclear physics for decades.
Elliott \cite{elliott} has demonstrated, that an average nuclear potential given by
a three dimensional harmonic oscillator corresponds to a SU(3) symmetry.  Low lying
collective states have been successfully described within the IBM-model \cite{arima}, which contains
as one limit the five dimensional harmonic oscillator, which is directly related to the Bohr-Mottelson
Hamiltonian. 

In this paper we will determine the symmetry group, which generates a single particle spectrum
similar to (\ref{nilsson}), but includes the magic numbers  right from the beginning.

Our approach is based on group theoretical methods developed
within the framework of  fractional calculus.

The fractional calculus \cite{f3}-\cite{he08} provides a set of axioms and methods to extend the coordinate and corresponding
derivative definitions in a reasonable way from integer order n to arbitrary order $\alpha$:
\begin{equation}
\{ x^n, {\partial^n \over \partial x^n} \} 
\rightarrow
\{ x^\alpha, {\partial^\alpha \over \partial x^\alpha} \}
\end{equation}
The definition of the fractional order derivative is not unique,  several definitions 
e.g. the Feller, Fourier, Riemann, Caputo, Weyl, Riesz, Gr\"unwald  fractional 
derivative definitions coexist \cite{f1}-\cite{grun}. A direct consequence of this diversity is the fact,
that the solutions e.g. of a one dimensional wave equation
differ significantly depending on the specific choice of a fractional derivative definition.

Until now it has always been assumed, that the fractional derivative type for an extension of a fractional
differential equation to multi-dimensional space  should be chosen uniquely. 

In contrast to this assumption, in this paper we will investigate 
properties of higher dimensional rotation groups  
with mixed Caputo and Riemann type definition of the fractional derivative.
We will demonstrate, that a  
fundamental dynamic symmetry is established, 
which determines the magic numbers for protons and neutron respectively  and furthermore 
 describes the ground state properties of nuclei with reasonable accuracy.  
\section{Notation}
We will investigate the spectrum of multi dimensional fractional rotation groups for two different definitions
of the fractional derivative, namely the Riemann- and Caputo fractional derivative. Both types are strongly
related.

Starting with the definition of the fractional Riemann integral 
\begin{equation}
\fl
{_\textrm{\tiny{R}}}I^\alpha \, f(x) = 
\cases{
({_\textrm{\tiny{R}}}I_{+}^\alpha f)(x) =  
\frac{1}{\Gamma(\alpha)}   
     \int_{0}^x  d\xi \, (x-\xi)^{\alpha-1} f(\xi)&\qquad \, $x \geq 0$\\
   ({_\textrm{\tiny{R}}}I_{-}^\alpha f)(x) =  
\frac{1}{\Gamma(\alpha)}   
     \int_x^0  d\xi \, (\xi-x)^{\alpha-1} f(\xi)&\qquad \,   $x<0$\\
}
\end{equation} 
where $\Gamma(z)$ denotes the Euler $\Gamma$-function, 
the fractional Riemann derivative is defined as the product of a fractional integration followed by an
ordinary differentiation:
\begin{equation}
\label{dr}
{_\textrm{\tiny{R}}}\partial_x^\alpha =  \frac{\partial}{\partial x}  {_\textrm{\tiny{R}}}I^{1-\alpha}
\end{equation} 
It is explicitely given by:
\begin{equation}
\label{driemann}
\fl
{_\textrm{\tiny{R}}}\partial_x^\alpha \, f(x) = 
\cases{
({_\textrm{\tiny{R}}}\partial_{+}^\alpha f)(x) =  
\frac{1}{\Gamma(1 -\alpha)} \frac{\partial}{\partial x}  
     \int_{0}^x  d\xi \, (x-\xi)^{-\alpha} f(\xi)&$x \geq 0$\\
   ({_\textrm{\tiny{R}}}\partial_{-}^\alpha f)(x) =  
\frac{1}{\Gamma(1 -\alpha)} \frac{\partial}{\partial x}  
     \int_x^0  d\xi \, (\xi-x)^{-\alpha} f(\xi)&$x<0$\\
}
\end{equation} 
The Caputo definition of a fractional derivative follows an inverted sequence of operations (\ref{dr}).
An ordinary differentiation is followed by a fractional integration
\begin{equation}
{_\textrm{\tiny{C}}}\partial_x^\alpha =   {_\textrm{\tiny{R}}}I^{1-\alpha} \frac{\partial}{\partial x} 
\end{equation} 
which results in:
\begin{equation}
\fl
{_\textrm{\tiny{C}}}\partial_x^\alpha \, f(x) = 
\cases{
({_\textrm{\tiny{C}}}\partial_{+}^\alpha f)(x) =  
\frac{1}{\Gamma(1 -\alpha)}   
     \int_{0}^x  d\xi \, (x-\xi)^{-\alpha} \frac{\partial}{\partial \xi}f(\xi)&$x \geq 0$\\
   ({_\textrm{\tiny{C}}}\partial_{-}^\alpha f)(x) =  
\frac{1}{\Gamma(1 -\alpha)}  
     \int_x^0  d\xi \, (\xi-x)^{-\alpha} \frac{\partial}{\partial \xi}f(\xi)&$x<0$\\
}
\end{equation} 
Applied to  a function set $f(x)=x^{n \alpha}$ using the Riemann fractional derivative definition (\ref{driemann}) we
obtain:
\begin{eqnarray}
{_\textrm{\tiny{R}}}\partial_x^\alpha \, x^{n \alpha}  &=& \frac{\Gamma(1+n \alpha)}{\Gamma(1+(n-1)\alpha)} \, x^{(n-1)\alpha}\\
\label{rx}
&=& {_\textrm{\tiny{R}}}[n]  \, x^{(n-1)\alpha}
\end{eqnarray} 
where we have introduced the abbreviation ${_\textrm{\tiny{R}}}[n]$.

For the Caputo definition of the fractional derivative it follows for the same function set:
\begin{eqnarray}
{_\textrm{\tiny{C}}}\partial_x^{\alpha} \, x^{n \alpha} &=& 
\cases{
\frac{\Gamma(1+n \alpha)}{\Gamma(1+(n-1)\alpha)} \, x^{(n-1)\alpha}&$n > 0$\\
0&$n=0$\\
}
\\
\label{cx}
&=& {_\textrm{\tiny{C}}}[n]  \, x^{(n-1)\alpha}
\end{eqnarray} 
where we have introduced the abbreviation $ {_\textrm{\tiny{C}}}[n]$. 

Both derivative definitions only differ in the case $n=0$: 
\begin{eqnarray}
{_\textrm{\tiny{C}}}[n]  &=&   {_\textrm{\tiny{R}}}[n] -\delta_{n0}\, {_\textrm{\tiny{R}}}[0]   \\
&=&  {_\textrm{\tiny{R}}}[n] - \delta_{n0}\,  \frac{1}{\Gamma(1-\alpha)}
\end{eqnarray} 
where $\delta_{mn}$ denotes the Kronecker-$\delta$.
We will rewrite  equations (\ref{rx}) and (\ref{cx}) simultaneously, introducing the short hand notation 
\begin{equation}
{_\textrm{\tiny{R,C}}}\partial_x^\alpha \, x^{n \alpha}  ={_\textrm{\tiny{R,C}}}[n]  \, x^{(n-1)\alpha}\\
\end{equation}
We now introduce the fractional angular momentum operators or generators of infinitesimal rotations 
in the $i,j$ plane on the $N$-dimensional Euclidean space:
 \begin{equation}
{_\textrm{\tiny{R,C}}}L_{ij}(\alpha)  =
{_\textrm{\tiny{R,C}}}\,i \hbar(x_i^\alpha \partial_j^\alpha-x_j^\alpha\partial_i^\alpha)
\end{equation}
The commutation relations of the fractional angular momentum operators are isomorph to the fractional 
extension of the rotational group $SO(N)$
\begin{eqnarray}
{_\textrm{\tiny{R,C}}} \, [L_{ij}(\alpha),L_{kl}(\alpha)] &=&{_\textrm{\tiny{R,C}}}\, i\hbar
{f_{ijkl}}^{mn}L_{mn}(\alpha) \\
& &  \qquad\qquad\qquad i,j,k,l,m,n=1,2,..,N \nonumber
\end{eqnarray}
with structure coefficients ${_\textrm{\tiny{R,C}}}{f_{ijkl}}^{mn}$. Their explicit form depends on the function set the fractional
angular momentum operators act on and on the fractional derivative type used. 

The Casimir-operators of the fractional rotation group ${_\textrm{\tiny{R}}}SO^\alpha(3)$ based on the
Riemann fractional derivative definition have been derived in {\cite{he07}} and 
for ${_\textrm{\tiny{C}}}SO^\alpha(3)$ based on the
Caputo fractional derivative definition are given in {\cite{he05}}. We summarize the major results:
\begin{figure}
\begin{center}
\includegraphics[bb=0 0 800 608,width=80mm,height=61mm]{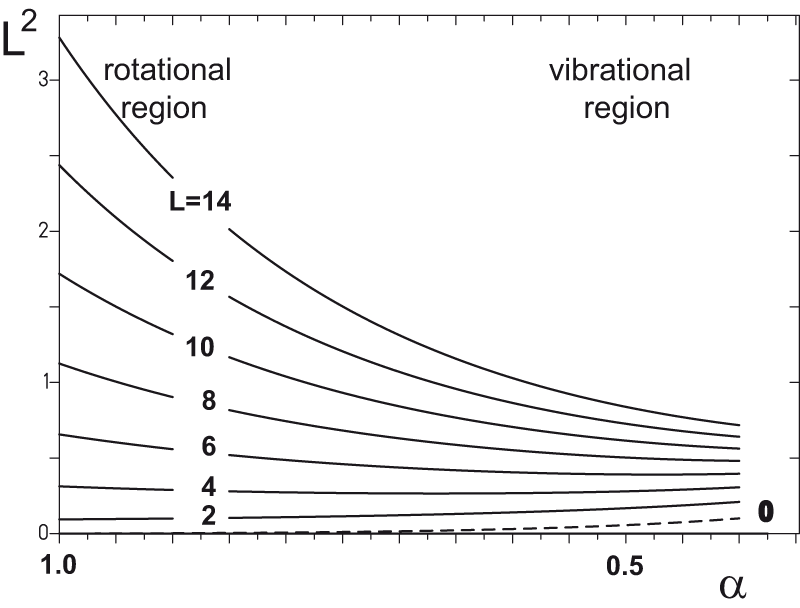}\\
\caption{\label{so3}
Spectrum of the Casimir operator $L^2(L,\alpha)$ from (\ref{L2}) as a function of the fractional derivative coefficient $\alpha$. Only
the $L=0$ state differs for Riemann and Caputo derivative.
} 
\end{center}
\end{figure}

According to the group chain
\begin{equation}
{_\textrm{\tiny{R,C}}}SO^\alpha(3) \supset {_\textrm{\tiny{R,C}}}SO^\alpha(2)
\end{equation}
there are two Casimir-operators $\Lambda_i$, namely $\Lambda_2 =L_z(\alpha)  = L_{12}(\alpha)$ and
$\Lambda_3  = L^2(\alpha) =  L_{12}^2 (\alpha) + L_{13}^2 (\alpha) + L_{23}^2 (\alpha)$. We 
introduce the two quantum numbers $L$ and $M$, which completely determine the eigenfunctions
$|LM>$.
It follows
\begin{eqnarray}
\label{Lz}
{_\textrm{\tiny{R,C}}} \, L_z(\alpha) |LM>  &=&
{_\textrm{\tiny{R,C}}} \, \hbar \, \textrm{sign}(M)\, [|M|]\, |LM> \\ 
 & & \qquad\qquad\qquad M = -L,-L+1,...,\pm 0,...,L \nonumber \\
\label{L2}
{_\textrm{\tiny{R,C}}} \, L^2(\alpha) |LM>  &=&
{_\textrm{\tiny{R,C}}} \, \hbar^2 [L][L+1] \,|LM> \\
 & & \qquad\qquad\qquad  L = 0,1,2,... \nonumber
\end{eqnarray}
where $|M|$ denotes the absolute value of $M$.
In addition, on the set of eigenfunctions $|LM>$, the parity operator $\Pi$ is diagonal and has the
eigenvalues
\begin{equation} 
\label{parity}
\Pi |LM> = (-1)^L |LM>
\end{equation} 
In figure \ref{so3} the eigenvalues of the Casimir-operator $L^2$ are shown as a function of $\alpha$. 
Only in  the case $L=0$ the spectra differ for the Riemann- and Caputo derivative.
While for the Caputo derivative 
\begin{equation}
{_\textrm{\tiny{C}}} \, L^2(\alpha) |00> = 0   
\end{equation}
because
${_\textrm{\tiny{C}}} [0] = 0 $, using the Riemann derivative for $\alpha \neq 1$ there is a nonvanishing contribution
\begin{equation} 
{_\textrm{\tiny{R}}} L^2(\alpha)|00> = {_\textrm{\tiny{R}}} \hbar^2 [0][1] |00> 
= \hbar^2 \frac{\Gamma(1+\alpha)}{\Gamma(1-\alpha)}|00>
\end{equation} 
\section{The Caputo-Riemann-Riemann symmetric rotor}
We now use group theoretical methods to construct higher dimensional representations of the fractional 
rotation groups ${_\textrm{\tiny{R,C}}}SO^\alpha(3)$. 

As an example of physical relevance we introduce the group ${_\textrm{\tiny{CRR}}}G$ with the following 
chain of subalgebras: 
\begin{equation}
{_\textrm{\tiny{CRR}}}G \supset {_\textrm{\tiny{C}}}SO^\alpha(3)
 \supset {_\textrm{\tiny{R}}}SO^\alpha(3)
 \supset {_\textrm{\tiny{R}}}SO^\alpha(3)
\end{equation}
The Hamiltonian $H$ can now be written in terms of the Casimir operators of the algebras appearing in the
chain and can be analytically diagonalized in the corresponding basis. The Hamiltonian is:
\begin{equation}
\label{hamilton}
H = \frac{\omega_1}{\hbar} {_\textrm{\tiny{C}}} L_1^2(\alpha)+
    \frac{\omega_2}{\hbar} {_\textrm{\tiny{R}}} L_2^2(\alpha)+
    \frac{\omega_3}{\hbar} {_\textrm{\tiny{R}}} L_3^2(\alpha) 
\end{equation}
with the free parameters $\omega_1,\omega_2,\omega_3$ and the basis is $|L_1 M_1 L_2 M_2 L_3 M_3>$.
Furthermore, we impose the following symmetries: 

First, the wave functions should be invariant under
parity transformations, which according to (\ref{parity}) leads to the conditions
\begin{equation}
\label{sym1}
L_1 = 2 n_1 \quad L_2 = 2 n_2 \quad L_3 = 2 n_3, \quad n_1,n_2,n_3=0,1,2,3,...
\end{equation}
second, we require
\begin{eqnarray}
\label{sym2}
{_\textrm{\tiny{C}}}L_{z_1}(\alpha)|L_1 M_1 L_2 M_2 L_3 M_3> &=& {_\textrm{\tiny{C}}}+\hbar[L_1]|L_1 M_1 L_2 M_2 L_3 M_3> \\
{_\textrm{\tiny{R}}}L_{z_2}(\alpha)|L_1 M_1 L_2 M_2 L_3 M_3> &=& {_\textrm{\tiny{R}}}+\hbar[L_2]|L_1 M_1 L_2 M_2 L_3 M_3> \\
{_\textrm{\tiny{R}}}L_{z_3}(\alpha)|L_1 M_1 L_2 M_2 L_3 M_3> &=& {_\textrm{\tiny{R}}}+\hbar[L_3]|L_1 M_1 L_2 M_2 L_3 M_3> 
\end{eqnarray}
which reduces the multiplicity of a given $|2n_1 M_1 2n_2 M_2 2n_3 M_3>$ set to 1.

With these conditions, the eigenvalues of the Hamiltonian (\ref{hamilton}) are given as
\begin{eqnarray}
\label{e1}
E(\alpha) &=& \hbar \omega_1 {_\textrm{\tiny{C}}} [2n_1][2n_1+1]+ \nonumber \\
& & 
      \hbar \omega_2 {_\textrm{\tiny{R}}} [2n_2][2n_2+1]+
      \hbar \omega_3 {_\textrm{\tiny{R}}} [2n_3][2n_3+1] \\
\label{e2}
&=& \sum_{i=1}^3 \hbar \omega_i \frac{\Gamma(1 + (2 n_i +1) \alpha)}{\Gamma(1 + (2 n_i-1) \alpha)}
- \delta_{n_1 0} \hbar \omega_1 \frac{\Gamma(1 +  \alpha)}{\Gamma(1 - \alpha)}\\
& &  \qquad\qquad\qquad\qquad\qquad n_1,n_2,n_3=0,1,2,.. \nonumber
\end{eqnarray}
on a basis $| 2 n_1 2 n_1 2 n_2 2 n_2 2 n_3 2 n_3>$.

This is the major result of our derivation. We call this model the Caputo-Riemann-Riemann symmetric rotor. 
What makes this model remarkable is its behaviour near $\alpha=1/2$. 
\begin{figure}
\begin{center}
\includegraphics[bb=0 0 800 550,width=120mm,height=82mm]{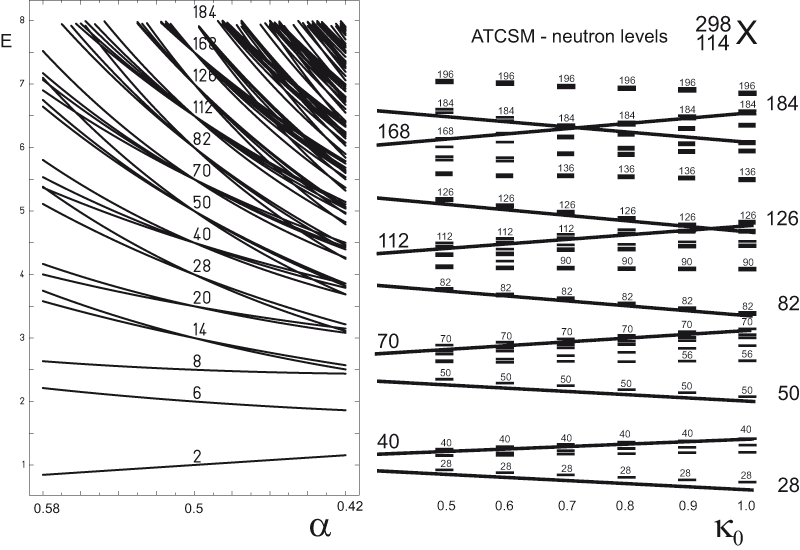}\\
\caption{\label{fig2}
On the left the
energy spectrum $E(\alpha)$ 
from (\ref{e2}) for the spherical case (\ref{sphere}) in units of $\hbar \omega_0$
for the Caputo-Riemann-Riemann
symmetric rotor is presented. The right diagram shows  the neutron energy levels for the spherical nucleus $^{298}_{114}X$ 
calculated within the framework of the asymmetric two center shell model (ATCSM) \cite{mah} near the ground state as
a function of increasing strength of the spin-orbit term ($\kappa_0 \kappa \vec{l} \vec{s}$) 
increasing from  $50\%$ to $100\%$ 
of the recommended
$\kappa$ value, while the $\mu l^2$ value is kept constant. 
The transition from magic numbers of the standard 3-dimensional harmonic oscillator levels
(\ref{set1}) to 
the shifted set of magic numbers (\ref{set2}) is  pointed out with thick lines. Left and right figure therefore
show a similar behaviour for the energy levels.  
} 
\end{center}
\end{figure}

On the left of figure \ref{fig2} we have plotted the energy levels in the vicinity of $\alpha \approx 1/2$
for the case 
\begin{equation}
\label{sphere}
\omega_1=\omega_2=\omega_3=\omega_0
\end{equation}
which we denote as the spherical case. 

For the idealized case $\alpha=1/2$, using the relation $\Gamma(1+z)= z\Gamma(z)$ 
the level spectrum (\ref{e2}) is simply given by:
\begin{equation}
\label{e12}
E(\alpha=1/2) = \hbar \omega_0 (  n_1 +n_2 +n_3 +\frac{3}{2} - \frac{1}{2}\delta_{n_1 0})  
\end{equation}
For $n_1 \neq 0$ this is the well known spectrum of the 3-dimensional harmonic oscillator. Assuming a twofold
spin degeneracy of the energy levels, we introduce the quantum number $N$ as
\begin{equation} 
N = n_1 + n_2 + n_3 
\end{equation} 
Consequently we obtain a first set $n_{\textrm{magic 1}} $ of magic numbers $n_{\textrm{magic}}$
\begin{eqnarray} 
\label{set1}
n_{\textrm{magic 1}} &=& \frac{1}{3}(N+1)(N+2)(N+3) \qquad N=1,2,3,...\\
&=& 8,20,40,70,112,168,240,...
\end{eqnarray} 
which correspond to the standard 3-dimensional harmonic oscillator at energies
\begin{equation} 
E = \hbar \omega_0 (N+3/2)
\end{equation} 
In addition, for $n_1 = 0$, which corresponds to the $|00 \, 2n_2 2n_2 \, 2n_3 2n_3>$ states,
we obtain a second set $n_{\textrm{magic 2}} $ of  magic numbers 
\begin{eqnarray} 
\label{set2}
n_{\textrm{magic 2}} &=& \frac{1}{3}(N+1)(5+(N+1)^2) \qquad N=0,1,2,3,...\\
&=& 2,6,14,28,50,82,126,184,258,...
\end{eqnarray} 
at energies
\begin{equation} 
E = \hbar \omega_0 (N+1)
\end{equation} 
which is shifted by the amount $- \frac{1}{2} \hbar \omega_0$ compared to the standard 3-dimensional 
harmonic oscillator values.

From figure \ref{fig2} it follows, that for $\alpha<1/2$ the second set $n_{\textrm{magic 2}}$ of energy levels falls off more rapidly than the levels of set
$n_{\textrm{magic 1}}$. As a consequence for decreasing $\alpha$ the magic numbers $n_{\textrm{magic 1}}$ die out
successively.  On the other hand, for $\alpha>1/2$ the same effect causes the magic numbers $n_{\textrm{magic 1}}$ to
survive.

We want to emphasize, that the described behaviour for the energy levels in the region $\alpha<1/2$ may be
directly compared to the influence of a $ls$-term in phenomenological shell models. As an example, on the
right hand side of figure \ref{fig2} a sequence of neutron levels for the superheavy element $^{298}_{114}X$ calculated
with the asymmetric two center shell model (ATCSM) \cite{mah} with increasing strength of the $ls$-term from
$50\%$ to $100\%$ is plotted. It shows, that the $n=168$ gap breaks down at about $70\%$ and the
$n=112$ gap at about $90\%$ of the recommended $\kappa$-value for the $ls$-term. This corresponds to 
an $\alpha \approx 0.46$ value, since in the Caputo-Riemann-Riemann symmetric rotor the $n=168$ gap breaks down
at $\alpha=0.466$, the $n=112$ gap at $\alpha=0.460$ and the $n=70$ gap vanishes at $\alpha=0.453$.  

We conclude, that the Caputo-Riemann-Riemann symmetric rotor predicts a well defined 
set of magic numbers. This set is a direct consequence of the underlying dynamic symmetries of the
three fractional rotation groups involved. It is indeed remarkable, that the same set of magic numbers 
is realized in nature as magic proton and neutron numbers. 

In the next section we will demonstrate, that the proposed analytical model is an appropriate tool to 
describe the ground state properties of nuclei.
\section{Ground state properties of nuclei }
We will use the Caputo-Riemann-Riemann symmetric rotor (\ref{e2}) as a dynamic shell model for a description
of the microscopic part of the total energy $E_{\textrm{tot}}$ of the nucleus.
\begin{eqnarray}
E_{\textrm{tot}} &=& E_{\textrm{macroscopic}}+E_{\textrm{microscopic}}\\
                 &=& E_{\textrm{macroscopic}}+\delta U + \delta P
\end{eqnarray}
where $\delta U$ and $\delta P$ denote the shell- and pairing energy contributions. 

For the macroscopic contribution we use the finite range liquid drop model 
(FRLDM) proposed by M\"oller \cite{mol} using the original parameters, except the value for the constant
energy contribution $a_0$. 

As the primary deformation parameter we use the ellipsoidal deformation $Q$:
\begin{equation}
Q = {b \over a} = {\omega_3 \over  \omega_1}= {\omega_3 \over  \omega_2}
\end{equation}
where $a,b$ are the
semi-axes of a rotational symmetric ellipsoid. Consequently a value $Q<1$ describes prolate and a value of
$Q>1$ descibes oblate shapes. 
In order to relate the ellipsiodal deformation $Q$ to the quadrupole deformation $\epsilon_2$ used by M\"oller, 
we define:
\begin{equation}
Q = 1-1.43085 \epsilon_2+0.707669 \epsilon_2^2
\end{equation}
Furthermore we extend the original FRLDM-model introducing an additional  curvature energy term $V_R(Q)$, 
which describes the interaction of the nucleus with the collective curved
coordinate space \cite{he81}:
\begin{equation} 
 V_R(Q) = -a_R B_R A^{-5/3} 
\end{equation} 
where $A$ is the nucleon number, $a_R$ is the curvature parameter given in $[{\textrm{MeV}}]$ and 
the relative curvature energy $B_R(Q)$ given as:
\begin{equation} 
B_R(Q)=  9\, Q^{16/3}\left( \frac{199-288 \log(2)}{(2+Q^2)(266-67 Q^2 + 96 (Q^2-4) \log(2))}\right)^2
\end{equation} 
which is normalized relative to a sphere $B_R(Q=1)=1$.

Therefore the total energy may be splitted into 
\begin{equation} 
E_{\textrm{tot}} = E_{\textrm{mac}}+E_{\textrm{mic}}
\end{equation} 
where
\begin{eqnarray}
\fl
E_{\textrm{mac}}(a_0,a_R)    &=& {\textrm{FRLDM}}(a_0,Q=1) + V_R(Q=1,a_R) \\
\fl
E_{\textrm{mic}}(a_0,a_R,Q) &=& +\delta U + \delta P + {\textrm{FRLDM}}(a_0,Q) + V_R(Q,a_R)  \nonumber \\
\fl
& & - \left( {\textrm{FRLDM}}(a_0,Q=1) + V_R(Q=1,a_R) \right)  
\end{eqnarray}
with two free parameters $a_0,a_R$, which will be used for a least square fit with the experimental data.

For calculation of
the shell corrections we use the Strutinsky method \cite{str1},\cite{str2}. 
Since we expect that the shell corrections are the
dominant contribution to the microscopic energy, for a first comparison with experimental data we 
will neglect the pairing energy term.  

\begin{figure}
\begin{center}
\includegraphics[bb=0 0 800 276,width=120mm,height=41mm]{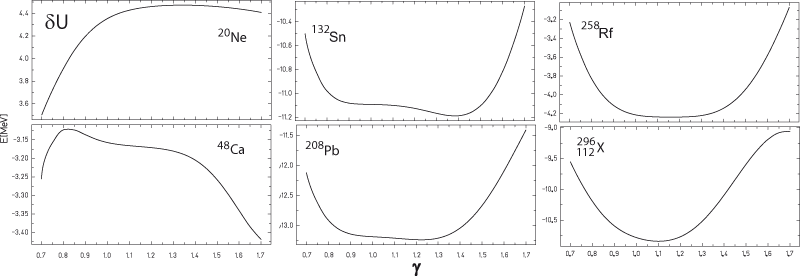}\\
\caption{\label{plateau}
As a test of the plateau condition $\partial U / \partial \gamma = 0$ for the Strutinsky shell correction method, the
total shell correction energy $\delta U = \delta U_P + \delta U_N$ is plotted as a function of $\gamma$ for
different nuclei. 
} 
\end{center}
\end{figure}

\begin{figure}
\begin{center}
\includegraphics[bb=0 0 800 570,width=120mm,height=86mm]{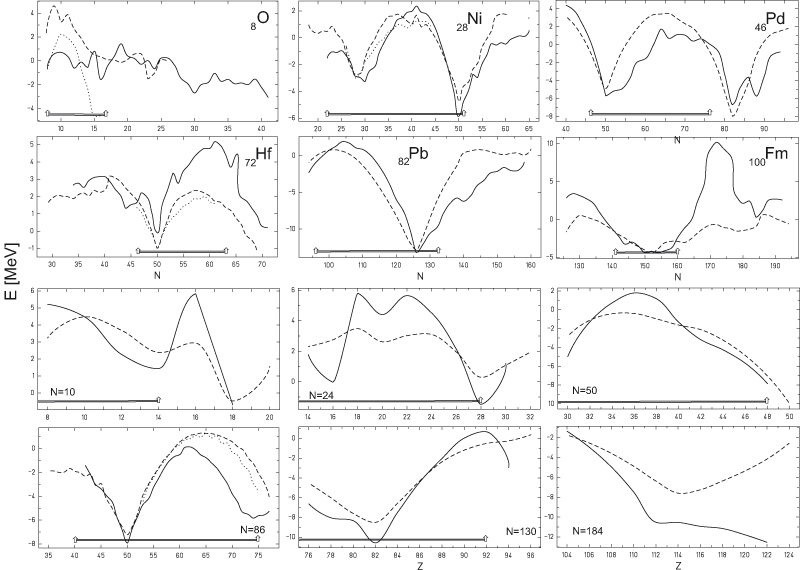}\\
\caption{\label{someshell}
Comparison of calculated shell corrections $\delta U$ from the Caputo-Riemann-Riemann symmetric rotor (\ref{e2}) with
the parameter set (\ref{om0})-(\ref{m}) (thick line) with the tabulated $E_{\textrm{mic}}$ from M\"oller \cite{mol} 
(dashed line). Upper two rows show values for a given $Z$ as a function of $N$, lower two rows for a given $N$ as
a function of $Z$. Bars indicate the experimentally known region. The original $\epsilon_2$ values from \cite{mol} are
used, which is the main source of error.  
} 
\end{center}
\end{figure}
\begin{figure}
\begin{center}
\includegraphics[bb=0 0 800 833,width=120mm,height=125mm]{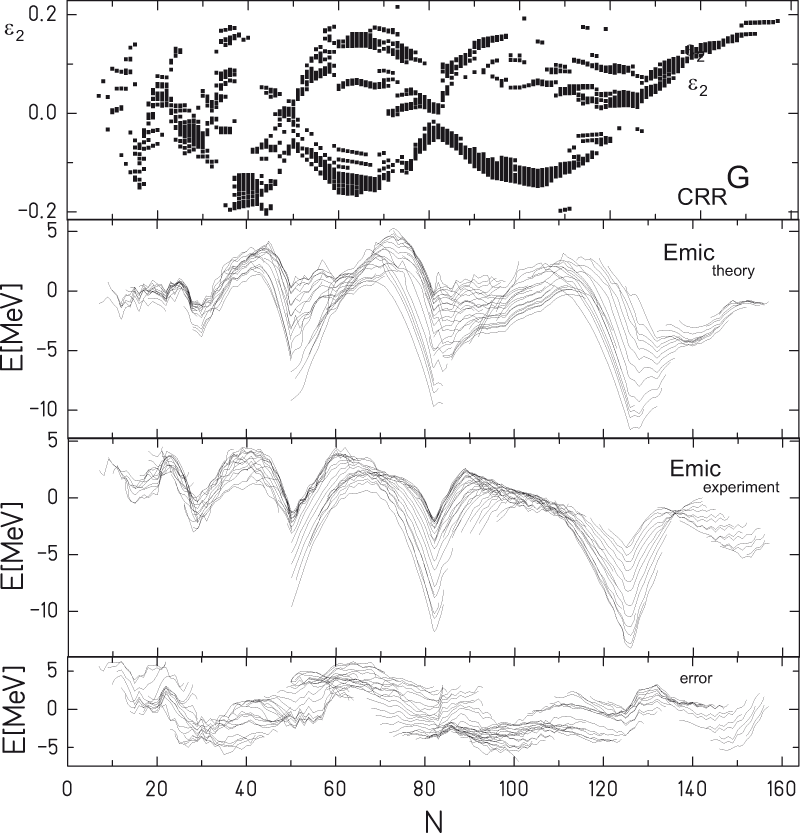}\\
\caption{\label{compare}
Comparison of calculated $E_{\textrm{mic}}$ from the Caputo-Riemann-Riemann symmetric rotor (\ref{e2}) with
the parameter set (\ref{om0})-(\ref{m}), minimized with respect to $\epsilon_2$  
with the experimental masses from Audi \cite{audi}
as a function of $N$. From top to bottom the minimized $\epsilon_2$ values, theoretical $E_{\textrm{mic}}$, 
experimental microscopic contribution from the difference of experimental mass excess and and macroscopic
FRLDM energy and error in $[MeV]$ are plotted. 
} 
\end{center}
\end{figure}

In order to calculate the shell corrections, we introduce the following parameters:    
\begin{eqnarray}
\label{om0}
\hbar \omega_0 &=& 38 A^{-\frac{1}{3}} [MeV]\\
\label{om1}
\omega_1 &= & \omega_0  Q^{-\frac{1}{3}}\\
\label{om2}
\omega_2 &= & \omega_1\\
\label{om3}
\omega_3 &= &\omega_0  Q^\frac{2}{3}\\
\label{az}
\alpha_Z &=& 
\cases{
0.46 + 0.000220 \, Z&$Z > 50$\\
0.2469 +0.00448 \, Z&$28<Z \leq 50$\\
0.2793 +0.00332 \, Z&$Z \leq 28$\\
}
\\
\label{an}
\alpha_N &=& 
\cases{
0.41 + 0.000200 \, N&$N > 50$\\
0.3118 +0.00216 \, N&$28<N \leq 50$\\
0.2793 +0.00332 \, N&$N \leq 28$\\
}
\\
\label{gam}
\gamma &=& 1.1 \, \hbar \omega_0\\
\label{m}
m&=& 4\\
a_0 &=& 2.409 [{\textrm{MeV}}]\\
a_R &=& 15.0 [{\textrm{MeV}}]
\end{eqnarray}
Input parameters are the number of protons $Z$, number of neutrons $N$, the nucleon number $A=N+Z$,  and
the ground state quadrupole deformation $\epsilon_2$. 

The values obtained include 
the frequencies
(\ref{om1}),(\ref{om2}),(\ref{om3}), which
result from a least square fit and quadratic approximation of equipotential surfaces, 
the fractional derivative coefficients for protons (\ref{az}) and neutrons (\ref{an}) which determine
the level spectrum for protons and neutrons for the proton and neutron part of the
shell correction energy respectively 
from a fit of the set of nuclids $^{56}_{28}\textrm{Ni}$, $^{100}_{50}\textrm{Sn}$, $^{132}_{50}\textrm{Sn}$, 
$^{208}_{82}\textrm{Pb}$ and from
the requirement, that the neutron shell correction for $^{100}_{50}\textrm{Sn}$ should amount about $-5.1 [MeV]$,
(\ref{gam}) from the plateau condition $\partial U / \partial \gamma = 0$ (see figure \ref{plateau}) and (\ref{m})
the order of included Hermite polynomials for the Strutinski shell correction method. Finally 
$\hbar \omega_0, a_0, a_R$ from a fit
of the experimental mass excess given in \cite{audi}.     

We compare our results with for the microscopic energy contribution $E_{\textrm{mic}}$ 
with data from M\"oller et. al. \cite{mol} and use their 
tabulated $\epsilon_2$ values.
They have not
only listed data for experimental masses but also predictions for regions, not yet confirmed by experiment.

In figure \ref{someshell} we compare the calculated $\delta U$ values with the tabulated $E_{\textrm{mic}}$,
which is justified for almost spherical shapes ($\epsilon_2 \approx 0$). The
results agree very well within the expected errors (which are estimated $\approx 2\,[MeV]$ for the pairing energy 
and $0.5\,[MeV]$ for $E_{\textrm{mic}}$), especially in the region of experimentally known nuclei.

A remarkable difference between the calculated shell correction and tabulated $E_{\textrm{mic}}$ 
from M\"oller occurs for superheavy elements ($N=184$, last picture in figure \ref{someshell}). 
While phenomenological shell models predict
a pronounced minimum in the shell correction energy for $Z=114$ \cite{my}-\cite{mos2} 
the situation is quite different for
the rotor model, where two magic shell closures at $Z=112$ and $Z=126$ are given, but the $Z=112$ shell closure
is not strong enough to produce a local minimum in the shell correction energy plot as a function of $Z$. 
Instead, between $Z=112$ and $Z=126$, there emerges a slightly falling energy plateau, 
which makes the full region promising candidates for stable, long-lived superheavy elements.

While this result contradicts predictions made with phenomenological shell models, it 
supports  recent results obtained with relativistic mean field models \cite{ben}, which predict a similar 
behaviour in the region of super heavy elements as the proposed  rotor model.
    
In figure \ref{compare} we have covered the complete region of available experimental data for nuclids 
and compare the calculated theoretical microscopic energy contribution minimized with respect to the deformation 
with the experimental values. The influence of shell closures is very clear. The rms-error is about 
$2.4 [{\textrm{MeV}}]$. The maximum deviation occures between
closed magic shells.  Therefore in the next section we will introduce a generalization of the 
proposed fractional rotor model, which not only determines the magic numbers accurately but in addition
determines the fine structure of the single particle spectrum correctly. 

\section{Fine structure of the single particle spectrum - the extended Caputo-Riemann-Riemann symmetric rotor}
In the previous section we have demonstrated, that the Caputo-Riemann-Riemann symmetric rotor correctly determines
the magic numbers in the single particle spectra for neutrons and protons. However, there
remains a significant difference between calculated and experimental ground state masses for nuclei with
nucleon numbers far from magic shell closures. This is a strong
indication for the fact, that the fine structure of the single particle levels is not yet correctly reproduced.

We therefore propose the following generalization of the Caputo-Riemann-Riemann symmetric rotor group: 
\begin{equation}
{_\textrm{\tiny{C3C2R3R3}}}G \supset {_\textrm{\tiny{C}}}SO^\alpha(3)
 \supset {_\textrm{\tiny{C}}}SO^\alpha(2)
 \supset {_\textrm{\tiny{R}}}SO^\alpha(3)
 \supset {_\textrm{\tiny{R}}}SO^\alpha(3)
\end{equation}
with the Casimir operators (\ref{Lz}) and (\ref{L2}) it follows for the 
Hamiltonian $H$:
\begin{equation}
\label{hamilton2}
H = \frac{\omega_1}{\hbar} {_\textrm{\tiny{C}}} L_1^2(\alpha)+
    B \omega_0 {_\textrm{\tiny{C}}} L_{z_1}(\alpha)+
    \frac{\omega_2}{\hbar} {_\textrm{\tiny{R}}} L_2^2(\alpha)+
    \frac{\omega_3}{\hbar} {_\textrm{\tiny{R}}} L_3^2(\alpha) 
\end{equation}
with the free parameters $\omega_1,\omega_2,\omega_3,B$, where $B$ may be called 
fractional magnetic field strength in units
$[\hbar \omega_0]$. 

Imposing the same  symmetries (\ref{sym1}),(\ref{sym2}) as in the case of the symmetric Caputo-Riemann-Riemann rotor, 
the eigenvalues of the Hamiltonian (\ref{hamilton2}) are given as
\begin{eqnarray}
\label{e111}
E(\alpha) &=& \hbar \omega_1 {_\textrm{\tiny{C}}} [2n_1][2n_1+1]+ 
B \hbar \omega_0 {_\textrm{\tiny{C}}} [2n_1]+ \nonumber \\
& & 
      \hbar \omega_2 {_\textrm{\tiny{R}}} [2n_2][2n_2+1]+
      \hbar \omega_3 {_\textrm{\tiny{R}}} [2n_3][2n_3+1] \\
\label{x2}
&=& \sum_{i=1}^3 \hbar \omega_i \frac{\Gamma(1 + (2 n_i +1) \alpha)}{\Gamma(1 + (2 n_i-1) \alpha)}
- \delta_{n_1 0} \hbar \omega_1 \frac{\Gamma(1 +  \alpha)}{\Gamma(1 - \alpha)}\nonumber\\
& & + B \hbar \omega_0   \frac{\Gamma(1 + (2 n_1) \alpha)}{\Gamma(1 + (2 n_1-1) \alpha)}
- \delta_{n_1 0} B \hbar \omega_0 \frac{1}{\Gamma(1 - \alpha)}\nonumber\\
& &  \qquad\qquad\qquad\qquad\qquad n_1,n_2,n_3=0,1,2,.. 
\end{eqnarray}
on a basis $| 2 n_1 2 n_1 2 n_2 2 n_2 2 n_3 2 n_3>$.

We call this model the extended Caputo-Riemann-Riemann symmetric rotor. The additional
${_\textrm{\tiny{C}}}L_{z_1}(\alpha)$ term  yields a level splitting of the 
harmonic oscillator set of magic numbers  $n_{\textrm{magic 1}}$ (\ref{set1}), 
while  the multiplicity of the $n_{\textrm{magic 2}}$  set (\ref{set2}) remains 
unchanged, since this set is characterized by $n_1=0$.  This is exactly the behaviour needed to describe 
the experimentally observed fine structure, as can be deduced from the right hand side of figure \ref{fig2}. 

In order to  clearly demonstrate  the influence of the additional term, we first investigate 
the level spectrum for the spherical (\ref{sphere}) and idealized case $\alpha=1/2$. 

The level spectrum (\ref{x2}) simply results as:
\begin{eqnarray}
\label{e33}
E(\alpha=1/2)& =& \hbar \omega_0 (  n_1 +n_2 +n_3 +\frac{3}{2} - \frac{1}{2}\delta_{n_1 0})\nonumber\\
& & 
+ B \hbar \omega_0  (\frac{n_1!}{\Gamma(1/2+n_1)} -  \frac{1}{\Gamma(1/2)}\delta_{n_1 0})  \\
   & =& \hbar \omega_0 (  n_1 +n_2 +n_3 +\frac{3}{2} - \frac{1}{2}\delta_{n_1 0}) \nonumber\\
& &
+ \frac{B \hbar \omega_0}{\sqrt{\pi}}   (\frac{2^{n_1} n_1!}{(2n_1-1)!!} -  \delta_{n_1 0})  
\end{eqnarray}
where $!!$ denotes the double factorial.

\begin{figure}
\begin{center}
\includegraphics[bb=0 0 800 421,width=120mm,height=63mm]{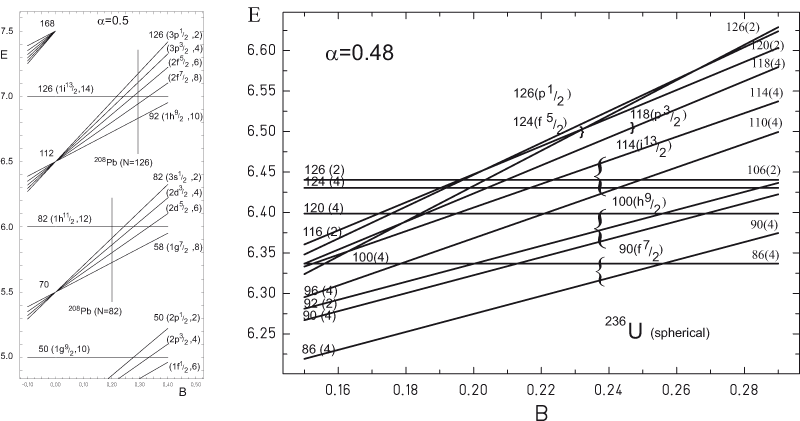}\\
\caption{\label{gc2}
For $\alpha=1/2$, on the left side the level spectrum
for the extended Caputo-Riemann-Riemann  symmetric rotor (\ref{x2}) is plotted as a function of the 
fractional magnetic field strength $B$. The levels are labeled
according to the corresponding $[Nlj_z]$ Nilsson scheme and the multiplicity is given. For $\alpha=0.48$ 
the resulting level sequence near $N \approx 126$ is plotted on the right. At $B \approx 0.25$ the resulting
spectrum coincides with the corresponding spherical Nilsson level spectrum. Brackets indicate
the proposed appropriate combinations of rotor levels. 
} 
\end{center}
\end{figure}
\begin{figure}
\begin{center}
\includegraphics[bb=0 0 800 828,width=120mm,height=124mm]{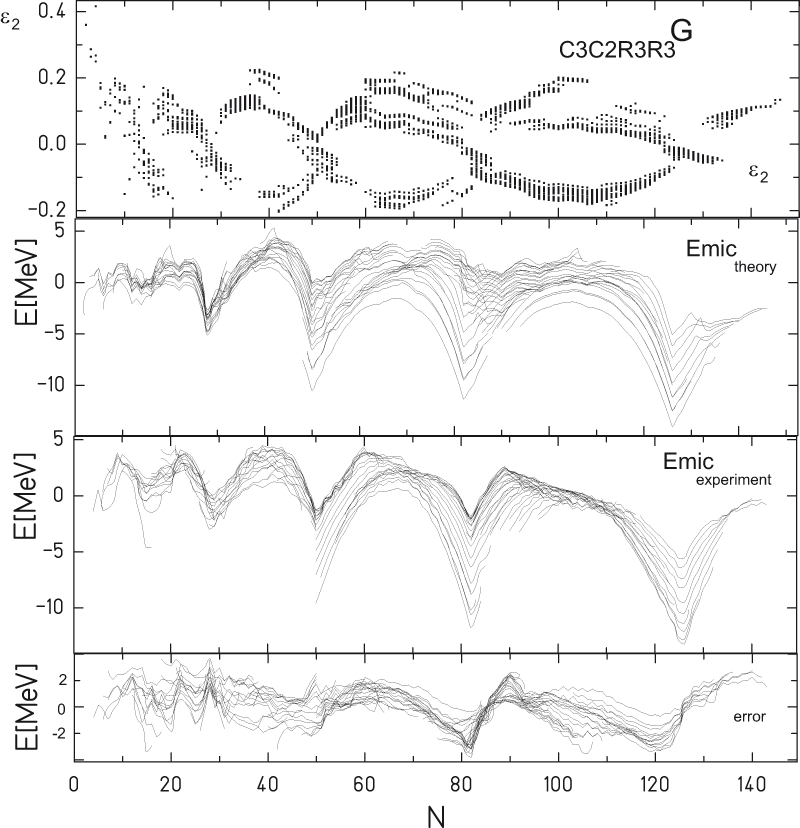}\\
\caption{\label{compare2}
Comparison of calculated $E_{\textrm{mic}}$ from the extended Caputo-Riemann-Riemann symmetric rotor (\ref{x2}) with
the parameter set (\ref{om02})-(\ref{BBB}), minimized with respect to $\epsilon_2$  with the experimental masses from Audi \cite{audi}
as a function of $N$. From top to bottom the minimized $\epsilon_2$ values, theoretical and experimental masses and
error in $[MeV]$ are plotted. 
} 
\end{center}
\end{figure}

On the left side of figure \ref{gc2} this spectrum is plotted in units $[\hbar \omega_0]$. Single levels
are labeled according to the Nilsson-scheme and multiplicities are given in brackets. For small fractional
field strength $B$ the resulting spectrum exactly follows the schematic level diagram of a phenomenological
shell model with spin-orbit term, as demonstrated e.g. by Goeppert-Mayer \cite{gop}. 

A small deviation from the ideal $\alpha = 1/2$ value reproduces the experimental spectra accurately: For
$\alpha=0.48$ the resulting level spectrum is given on the right hand side of figure \ref{gc2}. Obviously there
is an interference of two effects: First, for $\alpha \neq 1/2$ now the  degenerated levels of both magic sets
split up and second the fractional magnetic field $B$ acts on the subset $n_{\textrm{magic 1}}$.  For 
$B \approx 0.25$ the spectrum may be directly compared with the spherical Nilsson level scheme, which is given
for neutrons between $82 \leq N \leq 126$ as $2f{\frac{7}{2}}$, $1h{\frac{9}{2}}$,
$1i{\frac{13}{2}}$, $3p{\frac{3}{2}}$, $2f{\frac{5}{2}}$, $3p{\frac{1}{2}}$, see e.g. results of \cite{scharn}, 
which corresponds to a sequence of sub-shells at $90,100,114,118,124,126$. This sequence is correctly reproduced
with the extended Caputo-Riemann-Riemann symmetric rotor. 

With the parameter set, which is obtained by a fit with the experimental masses of Ca-, Sn- and Pb-isotopes
\begin{eqnarray}
\label{om02}
\hbar \omega_0 &=& 28 A^{-\frac{1}{3}} [MeV]\\
\label{az2}
\alpha_Z &=& 
\cases{
0.480 + 0.00022 \, Z&$Z > 50$\\
0.324 + 0.00332 \, Z&$Z \leq 50$\\
}
\\
\label{an2}
\alpha_N &=& 
\cases{
0.446 + 0.00022 \, N&$N > 29$\\
0.356 + 0.00332 \, N&$N \leq 29$\\
}
\\
\label{BBB}
B &=& 0.27 
\end{eqnarray}
the experimental masses are reproduced with an rms-error of $1.7 [{\textrm{MeV}}]$. Results are given in figure 
\ref{compare2}. 

The deformation parameters, obtained by minimization of the total energy, are to a large extend  consistent with
values given in \cite{mol} e.g. for $^{264}{\textrm{Hs}}_{108}$ we obtain $\epsilon_2 = 0.22$, which
conforms with  M\"oller`s ($\epsilon_2=0.2$) and Rutz`s results \cite{rutz}. However, there occur discrepancies
mostly for
exotic nuclei. For example our calculations determine the nucleus  $^{42}{\textrm{Si}}$ to be almost
spherical, while M\"oller predicts a definitely oblate shape. 

Finally, defining a nucleus with $\epsilon_2 > 0.05$ as prolate and with $\epsilon_2 < -0.05$ as oblate the
amount of prolate shapes is about $74 \%$ of all deformed nuclei. This is close to the  value of $82 \%$
\cite{taji}, obtained with the Nilsson model using the standard parameters.

Summarizing the results presented, the proposed 
extended Caputo-Riemann-Riemann symmetric rotor describes the ground state
properties of nuclei with reasonable accuracy. We  have demonstrated, that the nuclear shell structure may
indeed be successfully described on the basis of a dynamical symmetry model.

The advantages of this model, compared to phenomenological shell and relativistic mean field models 
respectively
are obvious:

Magic numbers are predicted, they are not the result of a fit with a phenomenological $ls$-term. There are no
potential-terms  or parametrized Skyrme-forces involved and finally, all results may be calculated analytically. 

The results obtained encourage further investigations in this field.  
The next steps  should include the pairing energy term  and parameters should be
determined by a more sophisticated fit procedure. With these additional contributions the model will most
probably describe nuclear properties with at least similar accuracy as the models currently used. 
\section{Conclusion}
Based on the Riemann- and Caputo definition of the fractional derivative we used the fractional
extensions of the standard rotation group $SO(3)$ to construct a higher dimensional representation
of a fractional rotation group with mixed derivative types. 

We obtained an extended symmetric rotor model, which predicts the sequence of magic proton and
neutron numbers accurately. Furthermore we have shown, that the ground state properties of
nuclei can be reproduced correctly  within the framework of this model.

Hence we have demonstrated, that a dynamic symmetry, generated by mixed fractional type rotation groups
is indeed realized in nature.

\section{Acknowledgment}
We thank A. Friedrich for useful discussions.

\end{document}